\nonstopmode

\newcommand{\Cal}[1]{{\cal #1}}
\newcommand{\ie}{i.e.{}}
\newcommand{\eg}{e.g.{}}
\newcommand{\etal}{\textit{et al.}}
\newcommand{\U}[1]{\,{\rm{#1}}}

\newcommand{\imag}{{\rm i}}
\newcommand{\euler}{\textrm{e}}
\newcommand{\mat}[1]{\hbox{\boldmath{$#1$}\unboldmath}}
\newcommand{\Sum}{\sum\limits}

\newcommand{\unitop}{\hat \mathbbm{1}}

\newcommand{\differential}{\>\textrm{d}}
\newcommand{\bra}[1]{\left<\right.\!#1\!\left.\right|}
\newcommand{\ket}[1]{\left|\right.\!#1\!\left.\right>}
\newcommand{\bracket}[2]{\left<\right.#1\>|\>#2\left.\right>}
\newcommand{\scalarprod}[3]{\bracket{#1}{#2\>|\>#3}}
\newcommand{\singlebracket}[4]{\left<\right.#1\,#2\>|\frac{1}%
{|\vec r_1 - \vec r_2|}|\>#3\,#4\left.\right>}

\newcommand{\Tr}{{\rm Tr}\,}

\newcommand{\offS}{S \hspace{-0.565em}\raisebox{0.23ex}{$\backslash$}}
\newcommand{\matOffS}{\hbox{\boldmath{$S\hspace{-0.64em}\raisebox{0.24ex}%
{$\backslash$}$}\unboldmath}}

\documentclass[prb,showpacs,nopreprintnumbers,twocolumn]{revtex4}
\usepackage{graphicx,subeqnarray,amsfonts,bbm,bm}
\usepackage[nativepdf,bookmarks,bookmarksopen,bookmarksnumbered,%
raiselinks,%
pdftitle={A priori Wannier functions from modified Hartree-Fock and Kohn-Sham equations},%
pdfauthor={Christian Buth},%
pdfsubject={Solid State Physics},%
pdfkeywords={ab initio, crystal orbital, Wannier function,%
localization}]{hyperref}

\begin{document}
\title{\emph{A priori} Wannier functions from modified Hartree-Fock and
Kohn-Sham equations}
\date{7~April 2007}
\author{Christian Buth}
\altaffiliation[Present address: ]{Argonne National Laboratory,
9700~South Cass Avenue, Argonne, Illinois~60439, USA}
\email[electronic mail: ]{Christian.Buth@web.de}
\affiliation{Max-Planck-Institut f\"ur Physik komplexer Systeme,
N\"othnitzer Stra\ss{}e~38, 01187~Dresden, Germany}

\begin{abstract}
The Hartree-Fock equations are modified to directly yield
Wannier functions following a proposal of
Shukla~\etal{} [Chem. Phys. Lett. \textbf{262}, 213--218 (1996)].
This approach circumvents the \emph{a posteriori}
application of the Wannier transformation to Bloch functions.
I give a novel and rigorous derivation of the relevant equations
by introducing an orthogonalizing
potential to ensure the orthogonality among the resulting
functions.
The properties of these, so-called \emph{a priori}
Wannier functions, are analyzed and
the relation of the modified Hartree-Fock
equations to the conventional, Bloch-function-based
equations is elucidated.
It is pointed out that the modified equations offer a different route to
maximally localized Wannier functions.
Their computational solution is found to involve an effort that is comparable
to the effort for the solution of the conventional equations.
Above all, I show how \emph{a priori} Wannier functions can be obtained
by a modification of the Kohn-Sham equations of density-functional theory.
\end{abstract}

%
%
%

\pacs{71.15.-m, 71.23.An, 31.15.Ne, 31.15.Pf}
\preprint{arXiv:cond-mat/0610356}
\maketitle

\section{Introduction}

Ever since the introduction of Wannier functions
in~1937,~\cite{Wannier:EE-37} they have been used
as an alternative representation to
Bloch functions in the study of crystalline
solids.~\cite{Wannier:EE-37,Blount:FB-62,Ashcroft:SSP-76,Callaway:QT-91,%
Fulde:EC-95,Marzari:ML-97,Zicovich:GM-01,Fulde:WF-02}
They provide a local, atomic-orbital-like view on
one-particle states of crystals, which is the direct
generalization to periodic systems of the concept
of localized molecular orbitals used by chemists
to study bonding in molecules and clusters.
In such a way, they come much closer to the intuitively
accessible concepts of chemical bonding than the conventional,
plane-wave-like Bloch orbitals which are completely
delocalized over the whole
crystal.~\cite{Wannier:EE-37,Blount:FB-62,Ashcroft:SSP-76,Callaway:QT-91,%
Fulde:EC-95,Marzari:ML-97,Zicovich:GM-01,Fulde:WF-02}
Wannier functions have been used frequently in formal deductions.
However, only recently do they become practically important, \eg,
they are used to achieve linear scaling with the system size of the
tight-binding or Kohn-Sham method.~\cite{Goedecker:LS-99,Wu:ON-02}
Moreover, as soon as electron correlations in the ground
state and in excited states are regarded, a local representation
offers big advantages.~\cite{Fulde:EC-95,Shukla:WFB-99,%
Fulde:WF-02,Albrecht:LAI-02,Buth:BS-04,Buth:MC-05,%
Buth:AG-05,Buth:HB-06,Buth:BHF-06,Pisani:MP-06}

Conventionally, Wannier functions are determined \emph{a posteriori}
from Bloch functions utilizing the Wannier
transformation.~\cite{Wannier:EE-37,Blount:FB-62,Ashcroft:SSP-76,%
Callaway:QT-91,Marzari:ML-97,Zicovich:GM-01}
It is unique up to a unitary matrix, which
can be chosen freely to obtain a certain set of Wannier functions,
\eg, those which are maximally localized with respect to a given criterion.
Marzari and Vanderbilt~\cite{Marzari:ML-97} and
Zicovich-Wilson~\etal~\cite{Zicovich:GM-01} advocate the Foster-Boys
functional.~\cite{Foster:MO-60,Boys:MO-60,Boys:-66}
However, there are a number of other localization
criteria such as the method of Edmiston and
Ruedenberg~\cite{Edmiston:LA-63} or the recipe of Pipek and
Mezey~\cite{Pipek:FI-89} which can be used, too.

Instead, Shukla~\etal~\cite{Shukla:EC-96} proposed an embedded cluster model
that facilitates solving modified Hartree-Fock equations, which directly
yield Wannier-type functions.
The theory has been implemented in terms of the computer program
\textsc{wannier},~\cite{Shukla:EC-96,Shukla:WF-98} which
has been used in a series of studies of polymers and
crystals.~\cite{Shukla:EC-96,Shukla:WF-98,Shukla:TC-98,%
Albrecht:HF-98,Shukla:WFB-99}
It was found, empirically, to provide accurate Hartree-Fock
energies and band structures.~\cite{Shukla:EC-96,Shukla:WF-98,Albrecht:HF-98,%
Abdurahman:AI-99}
The program has also been used as a starting point for
post Hartree-Fock calculations;
the correlation energy of the ground state~\cite{Shukla:WFB-99,Abdurahman:AI-00}
and the quasiparticle band structure of several polymers and crystals have been
investigated.~\cite{Albrecht:LAI-02,Buth:MC-05,Buth:AG-05,Buth:BHF-06}
The idea to directly solve for Wannier-type functions has also been
discussed, \eg, in Refs.~\onlinecite{Mauri:EC-94,%
Goedecker:LS-99,Wu:ON-02} (and references therein).

The paper is structured as follows.
In Sec.~\ref{sec:apriori}, I devise a rigorous proof of the modified
Hartree-Fock equations of Shukla~\etal~\cite{Shukla:EC-96} and give
the corresponding modified Roothaan-Hall equations.
The functions that result from these equations are termed \emph{a priori}
Wannier functions.
The relation of the modified equations to the canonical, Bloch-function-based
equations and a comparison of the computational effort of both approaches
can be found in Sec.~\ref{sec:properties}.
Conclusions are drawn in Sec.~\ref{sec:conclusion};
specifically, I show that the Kohn-Sham equations of density-functional
theory can be modified in the same way as the Hartree-Fock equations
to obtain \emph{a priori} Wannier functions.

Atomic units are used throughout the article.

\section{\emph{A priori} Wannier functions}
\label{sec:apriori}
\subsection{Definitions}

The sets of functions~$\{ w_{\vec R \, \varrho}(\vec r \, s),
\varrho = 1, \ldots, K \}$ are associated with each unit cell~$\vec R$
of a crystal;
they depend on the spatial and spin coordinates~$\vec r$ and $s$,
respectively.
The sets are translationally related, \ie, two such sets can
be brought into coincidence by displacing the functions in one of
them by a suitable lattice vector~$\vec R'$,
\begin{eqnarray}
  \label{eq:WannierTrans}
  w_{\vec R + \vec R' \, \varrho}(\vec r \, s) &\equiv&
  \bracket{\vec r \, s}{\vec R + \vec R' \, {\varrho}}
  = w_{\vec R \, \varrho}(\vec r - \vec R' \, s) \nonumber \\
  &=& \hat T_{\vec R'} w_{\vec R \, \varrho}(\vec r \, s) \; .
\end{eqnarray}
Here, $\hat T_{\vec R'}$ denotes the translation operator for a passive
translation by~$\vec R'$.
Furthermore, the functions are assumed to be orthonormal with
respect to an integration over the entire space,
\begin{eqnarray}
  \bracket{w_{\vec R \, \sigma}}{w_{\vec R' \, \varrho}}
  &=& \Sum_{s = -\frac{1}{2}}^{\frac{1}{2}} \int_{\mathbb R^3}
  w_{\vec R \, \sigma}^*(\vec r \, s) \> w_{\vec R' \, \varrho}(\vec r \, s)
  \differential^3 r \nonumber \\
  \label{eq:WannierOrth}
  &=& \delta_{\vec R \, \vec R'} \> \delta_{\sigma \, \varrho} \; .
\end{eqnarray}
The two properties~(\ref{eq:WannierTrans}) and (\ref{eq:WannierOrth})
characterize (spin) Wannier functions,~\cite{Wannier:EE-37,Blount:FB-62,%
Ashcroft:SSP-76,Callaway:QT-91} which are frequently termed
(spin) Wannier orbitals, if they are Hartree-Fock or Kohn-Sham (spin) orbitals.

I describe the crystal by a nonrelativistic
Hamiltonian~\cite{Fulde:EC-95,Ladik:PS-99}
\begin{subeqnarray}
  \label{eq:Operator_Hamiltonian}
  \slabel{eq:two-ham}
  \hat H &=& \Sum_{n=1}^N \hat h_n
    + \frac{1}{2} \, \Sum_{m,n=1 \atop m \neq n}^N \frac{1}
    {|\vec r_m - \vec r_n|}
    + \hat E_{\rm nucl} \; , \\
  \slabel{eq:one-ham}
  \hat h_n &=& - \frac{1}{2} \vec\nabla^2_n
    - \Sum_{\vec R} \Sum_{A=1}^M \frac{Z_{\vec R \, A}}
    {|\vec r_n - \vec r_{\vec R \, A}|} \; , \\
  \slabel{eq:nucl-ham}
  \hat E_{\rm nucl} &=& \frac{1}{2} \, {\textstyle
    \Sum_{\scriptstyle \vec R, \vec R' \atop \scriptstyle \vec R \neq \vec R'
    \lor A \neq B \hspace{-1cm}} \Sum_{A, B = 1}^M}
    \frac{Z_{\vec R \, A} Z_{\vec R' \, B}}
    {|\vec r_{\vec R \, A} - \vec r_{\vec R' \, B}|} \; ,
\end{subeqnarray}
assuming fixed nuclei.
Here, $N$~denotes the number of electrons in the crystal which is
represented by a parallelepiped that consists of $N_0$~unit cells.
The distance between the $m$th and the $n$th electron is represented
by~$|\vec r_m - \vec r_n|$.
The number of nuclei per unit cell is indicated by~$M$ and
$Z_{\vec R \, A} \equiv Z_A$~stands for the charge of
nucleus~$A$ in unit cell~$\vec R$.
Then, $|\vec r_n - \vec r_{\vec R \, A}|$~is the distance between the $n$th
electron and the $A$th nucleus in unit cell~$\vec R$.
Finally, $|\vec r_{\vec R \, A} - \vec r_{\vec R' \, B}|$~denotes the
distance between nuclei~$A$ and $B$ of charge~$Z_{\vec R \, A}
\equiv Z_A$ and $Z_{\vec R' \, B} \equiv Z_B$ in unit cells~$\vec R$
and $\vec R'$, respectively.

\subsection{Wannier-Hartree-Fock equations}

The ansatz for the Hartree-Fock wave function is given by the
Slater determinant in terms of the occupied Wannier orbitals,
\begin{widetext}
\begin{equation}
  \label{eq:WannierSlaterDet}
  \Phi^N_0(\vec r_1 \, s_1, \ldots, \vec r_N \, s_N) = \hat A \,
    \prod_{i=1}^{N_0} \prod_{j=1}^{N'} w_{\vec R_i \, \kappa_j}
    (\vec r_{j + (i-1) \, N'} \, s_{j + (i-1) \, N'}) \; ,
\end{equation}
where~$N' = \frac{N}{N_0}$ is the number of occupied
orbitals per unit cell and $\hat A$ stands for
the antisymmetrizer.~\cite{McWeeny:MQM-92}
The energy expectation value of the Slater
determinant~(\ref{eq:WannierSlaterDet}) with the
Hamiltonian~(\ref{eq:Operator_Hamiltonian}) reads~\cite{Szabo:MQC-89,%
McWeeny:MQM-92,Shukla:EC-96}
\begin{equation}
  \label{eq:WannierFockEnergy}
  \begin{array}{rcl}
    \hspace{-0.5em} E[\Phi^N_0(\vec r_1 \, s_1, \ldots, \vec r_N \, s_N)]
    &=& \Sum_{\vec R} \Sum_{\kappa = 1}^{N'}
      \> \scalarprod{w_{\vec R \, \kappa}}{\hat h_1}{w_{\vec R \, \kappa}}
      + \frac{1}{2} \Sum_{\vec R, \vec R'}
      \Sum_{\kappa, \xi = 1}^{N'}
      \Bigl[ \singlebracket{w_{\vec R \, \kappa}(1)}{w_{\vec R' \, \xi}(2)}
                           {w_{\vec R \, \kappa}(1)}{w_{\vec R' \, \xi}(2)} \\
      &&{} - \singlebracket{w_{\vec R \, \kappa}(1)}{w_{\vec R' \, \xi}(2)}
                           {w_{\vec R' \, \xi}(1)}{w_{\vec R \, \kappa}(2)}
      \Bigl] + E_{\rm nucl} \; ;
  \end{array}
\end{equation}
\end{widetext}
it is a functional of the Wannier orbitals.

In order to minimize the energy expectation value to obtain the best
description of the ground state by a single Slater determinant
in terms of Ritz' variational principle,
functional variation with respect to the orbitals in
Eq.~(\ref{eq:WannierFockEnergy}) is carried
out.~\cite{Szabo:MQC-89,McWeeny:MQM-92,Buth:MC-05}
In doing so, I subtract the constraints
\begin{equation}
  \label{eq:LagrangianMult}
    \Sum_{\vec R, \vec R'} \Sum_{\kappa, \xi = 1}^{N'}
      \Lambda_{\vec R' \, \xi \; \vec R \, \kappa}
      \, (\bracket{w_{\vec R \, \kappa}}{w_{\vec R' \, \xi}}
      - \delta_{\vec R \vec R'} \, \delta_{\kappa \xi})
\end{equation}
from~$E[\Phi^N_0(\vec r_1 \, s_1, \ldots, \vec r_N \, s_N)]$, employing
the Lagrangian multipliers~$\Lambda_{\vec R' \, \xi \;
\vec R \, \kappa}$;
this ensures the orthonormality of the orbitals.
One arrives at the Hartree-Fock equations in terms of Wannier orbitals,
\begin{equation}
  \label{eq:WannierFock}
  \begin{array}{rcl}
    \hat f \ket{w_{\vec R \, \kappa}}
      &=& \Sum_{\xi = 1}^{N'} \Lambda_{\vec R \, \xi \; \vec R \, \kappa} \,
      \ket{w_{\vec R \, \xi}} \\
      &&{} + \Sum_{\vec R' \neq \vec R} \Sum_{\xi = 1}^{N'}
      \Lambda_{\vec R' \, \xi \; \vec R \, \kappa} \,
      \ket{w_{\vec R' \, \xi}} \; ,
  \end{array}
\end{equation}
with $\hat f$ denoting the Fock operator.
Expression~(\ref{eq:WannierFock}) forms a set of $N$~equations
which couple the occupied Wannier orbitals in a unit cell to the
occupied Wannier orbitals in all other cells of the crystal.
Above all, the Hartree-Fock equations~(\ref{eq:WannierFock})
do not have the form of an eigenvalue equation.
For these reasons, their practical application is cumbersome.

To make progress towards a more favorable, modified form of Hartree-Fock
equations for Wannier orbitals, I omit the intercell Lagrangian
multipliers in Eq.~(\ref{eq:LagrangianMult}) and thus arrive at
the new energy functional
\begin{equation}
  \label{eq:CellLagrangianMult}
  \begin{array}{rr}
    &\Cal L[\Phi^N_0(\vec r_1 \, s_1, \ldots, \vec r_N \, s_N)] =
      E[\Phi^N_0(\vec r_1 \, s_1, \ldots, \vec r_N \, s_N)] \\
    &{} -\Sum_{\vec R} \Sum_{\kappa, \xi = 1}^{N'}
      \Lambda_{\xi \kappa} \, (\bracket{w_{\vec R \, \kappa}}
    {w_{\vec R \, \xi}} - \delta_{\kappa \, \xi}) \; .
  \end{array}
\end{equation}
Here, the translational symmetry of~$\mat \Lambda$ is exploited,
\ie, $\Lambda_{\xi \kappa} \equiv \Lambda_{\vec 0 \, \xi \; \vec 0 \, \kappa}
= \Lambda_{\vec R \, \xi \; \vec R \, \kappa}$
for all lattice vectors~$\vec R$ and orbital indices~$\xi$, $\kappa$.
This symmetry, however, will be broken again in the next paragraph.
Minimizing $\Cal L$ leads to orthonormal spin-orbitals in each
unit cell.
However, by this simplification of Eq.~(\ref{eq:LagrangianMult}),
one does not enforce the mutual intercell orthogonality of the
orbitals.
Starting from a properly orthonormalized initial guess for the Wannier
orbitals in the origin cell may lead to overlapping orbitals.

In order to nonetheless achieve orthogonality of the occupied orbitals in the
origin cell to all other orbitals, \ie, their periodic images in all the
other unit cells, I modify the energy functional~$\Cal L$
in Eq.~(\ref{eq:CellLagrangianMult}) once more by adding an artificial
orthogonalizing potential,
\begin{equation}
  \label{eq:WannierRitz}
  \begin{array}{lr}
    \displaystyle \Cal L'[\Phi^N_0(\vec r_1 \, s_1, \ldots, \vec r_N \, s_N)]
    = & \\
    & \displaystyle  \hspace{-2cm} \Cal L[\Phi^N_0(\vec r_1 \, s_1, \ldots,
    \vec r_N \, s_N)] + V_{\rm Orth} \; ,
  \end{array}
\end{equation}
which is defined by
\begin{equation}
  \label{eq:orthogpotdef}
  V_{\rm Orth} = \frac{\lambda}{2} \,
  \Sum_{\vec R, \vec R' \atop \vec R \neq \vec R'}
  \Sum_{\kappa, \xi = 1}^{N^\prime}
  \bracket{w_{\vec R          \, \kappa}}
          {w_{\vec R' \, \xi   }}
  \bracket{w_{\vec R' \, \xi   }}
          {w_{\vec R          \, \kappa}} \, ,
\end{equation}
where $\lambda > 0$~is called the orthogonalizing potential strength
or shift parameter.
All terms in definition~(\ref{eq:orthogpotdef}) are real and
non-negative, causing an increase of energy proportional to the
square of the modulus of the overlap between a pair of orbitals
in different unit cells~$\vec R$ and $\vec R'$.
Minimizing $\Cal L'$, in the limit~$\lambda \to \infty$, both preserves the
orthogonality among the Wannier orbitals in all cells of the
crystal and minimizes the Hartree-Fock energy functional~$E$
in Eq.~(\ref{eq:WannierFockEnergy}).
It has been shown in practical computations that a finite
orthogonalization potential strength~$\lambda$ in the
range of~$10^3$--$10^5 \U{hartrees}$ causes the resulting Hartree-Fock energies
not to show a noticeable dependence on~$\lambda$.~\cite{Shukla:TC-98}

Expression~(\ref{eq:orthogpotdef}) can be rewritten compactly by recognizing that
the orbitals within a particular unit cell are orthonormal by
construction~(\ref{eq:CellLagrangianMult}).
Therefore, the constraint~$\vec R \, \kappa \neq \vec R' \, \xi$
can be used instead for the summation in Eq.~(\ref{eq:orthogpotdef}).
Consequently, the potential~$V_{\rm Orth}$ can be expressed in terms of the
off-diagonal elements of the overlap matrix between the orbitals,
\begin{equation}
  \offS_{\vec R \, \kappa \; \vec R' \, \xi} = (1 - \delta_{\vec R \, \vec R'} \,
  \delta_{\kappa \, \xi}) \, \bracket{w_{\vec R \, \kappa}}{w_{\vec R' \, \xi}} \; .
\end{equation}
By taking the trace of~$\matOffS^2$, I arrive at
\begin{equation}
  \label{eq:orthogpot}
  V_{\rm Orth} = \frac{\lambda}{2} \, \Tr \matOffS^2 \; ,
\end{equation}
which is an alternate form for Eq.~(\ref{eq:orthogpotdef}).

To minimize the functional~$\Cal L'$ in Eq.~(\ref{eq:WannierRitz}),
I carry out functional variation with respect to the $N'$~occupied
orbitals in unit cell~$\vec R$;%
\footnote{Only orthogonal variations of the orbitals are considered here
which are guaranteed by using the technique of Lagrangian multipliers
in conjunction with the orthogonalizing potential.
Note, however, that, following the derivation of Adams~\cite{Adams:SH-61}
of Hartree-Fock equations for overlapping orbitals,
I find---setting in the end the initial overlap
matrix to unity---that the same equations are
obtained as for variations which maintain
the orbitals purely orthogonal.}
I arrive at modified Hartree-Fock equations
\begin{equation}
  \label{eq:WannierFockPenalty}
  (\hat f + \lambda \, \hat{\Cal P}_{\vec R}) \ket{w_{\vec R \, \kappa}}
  = \Sum_{\xi = 1}^{N^\prime} \Lambda_{\vec R \, \xi \; \vec R \, \kappa} \,
    \ket{w_{\vec R \, \xi}} \; ,
\end{equation}
which I term \emph{Wannier-Hartree-Fock equations}.
The penalty projection operator
\footnote{Note that Eq.~(\ref{eq:penalty}) only describes a projection
operator for orthonormal orbitals~$w_{\vec R' \, \xi}(\vec r \, s)$.}
therein is defined by
\begin{equation}
  \label{eq:penalty}
  \hat{\Cal P}_{\vec R} = \Sum_{\vec R' \neq \vec R} \Sum_{\xi=1}^{N^\prime}
  \ket{w_{\vec R' \, \xi}}\bra{w_{\vec R' \, \xi}} \; .
\end{equation}
It is not translationally symmetric;
instead, the relation~$\hat T_{\vec g} \, \hat{\Cal P}_{\vec R}
= \hat{\Cal P}_{\vec R - \vec g}$ holds.
This property of the projector~$\hat{\Cal P}_{\vec R}$
distinguishes the Wannier orbitals in unit cell~$\vec R$
in the Wannier-Hartree-Fock equations~(\ref{eq:WannierFockPenalty})
from their periodic images in other unit cells because
it breaks the translational symmetry of the Fock operator~$\hat f$.

The translational relation of the Wannier function~(\ref{eq:WannierTrans})
implies that it is sufficient to formulate and solve
Eq.~(\ref{eq:WannierFockPenalty}) only in the origin cell.
For orthogonal orbitals, the Lagrangian
multipliers~$\Lambda_{\vec 0 \, \kappa \; \vec 0 \, \xi}$
constitute a Hermitian matrix,~\cite{Szabo:MQC-89,McWeeny:MQM-92} which
is diagonalizable by a unitary transformation~$\mat{X^{\dagger}}
\mat\Lambda \mat X = \mat \varepsilon$.
Both $\hat f$ and $\hat{\Cal P}_{\vec 0}$ are invariant under such a
transformation which mixes the orbitals within every unit cell
including, particularly, the origin cell.
Therefore, I formally get a Hermitian $N^\prime \times
N^\prime$~eigenvalue problem
\begin{equation}
  \label{eq:WannierFockEigenvalue}
  (\hat f + \lambda \, \hat{\Cal P}_{\vec 0}) \ket{\check w_{\vec 0 \, \kappa}}
    = \varepsilon_{\vec 0 \, \kappa} \, \ket{\check w_{\vec 0 \, \kappa}} \; ,
\end{equation}
which resembles the canonical Hartree-Fock equations~\cite{Re:SC-67,%
Andre:ET-67,Szabo:MQC-89,McWeeny:MQM-92,Ladik:PS-99} and
is thus named pseudocanonical Wannier-Hartree-Fock equations.
The transformation of the orbitals is indicated by affixing a check accent.
They are referred to as pseudocanonical Wannier orbitals.
Due to the fact that these orbitals diagonalize the Fock matrix in
the origin cell, they are uniquely determined, apart
from degeneracies, in analogy to canonical
orbitals.~\cite{Szabo:MQC-89,McWeeny:MQM-92}
Once a self-consistent solution of Eq.~(\ref{eq:WannierFockEigenvalue})
has been found, the parametrical dependence on~$\lambda$ of the orbitals and
eigenvalues therein
vanishes because they are equal to the orthonormal
orbitals from the previous iteration that have been used to
construct~$\hat f$ and $\hat{\Cal P}_{\vec 0}$ to begin with.

The pseudocanonical Wannier orbitals are
delocalized over the entire origin cell and thus implicate a
similar disadvantageous nonlocality associated with Bloch
orbitals when applying cutoff criteria to the Fock matrix and
to the two-electron integrals.~\cite{Buth:MC-05,Buth:AG-05,Buth:BHF-06}
However, this form of the Wannier-Hartree-Fock equations is a good
starting point for further
improvements by means of an additional localizing potential that
can be introduced in the expression for~$\Cal L'$ in Eq.~(\ref{eq:WannierRitz}).
Established forms of localizing potentials are the one of
Edmiston and Ruedenberg~\cite{Edmiston:LA-63} for their
localization criterion and the one of Gilbert~\cite{Gilbert:SC-64}
for the Foster-Boys criterion.~\cite{Foster:MO-60,Boys:MO-60,Boys:-66}

The functional dependence of~$\hat f + \lambda \, \hat{\Cal P}_{\vec 0}$
on the occupied Wannier orbitals can be
disregarded.~\cite{Szabo:MQC-89,McWeeny:MQM-92}
Then, the Fock operator becomes a conventional Hermitian operator
and the restriction of the penalty projection operator,
to act only on occupied orbitals, can be released,
\ie, $N'$ in Eq.~(\ref{eq:penalty}) is replaced by the
total number of orbitals per unit cell~$K$.
Now, Eq.~(\ref{eq:WannierFockEigenvalue}) holds also for
virtual Wannier orbitals.

\subsection{Wannier-Roothaan-Hall equations}

Spin Wannier orbitals have been employed so far.
Let me assume a restricted, closed-shell Hartree-Fock point of view to
remove the spin dependence.~\cite{Szabo:MQC-89,McWeeny:MQM-92}
In this case, the spin orbitals are expressed
as the product of a spatial orbital~$\check w_{\vec R \,
\varrho}(\vec r)$ with the spinor for spin up~$\alpha(s)$
and spin down~$\beta(s)$, respectively.
The spatial orbitals, $\check w_{\vec R \,
\varrho}(\vec r)$, $\varrho = 1, \ldots, K$, are expanded in terms
of one-particle basis functions,~\cite{Shukla:EC-96,%
Zicovich:GM-01} $\chi_{\mu}(\vec r)$, $\mu = 1, \ldots, K$,
\begin{equation}
  \label{eq:Wannierbasis}
  \begin{array}{rcl}
    \check w_{\vec R \, \varrho}(\vec r) &=& \hat T_{\vec R} \> \Sum_{\vec g}
      \Sum_{\mu=1}^K C_{\vec g \, \mu \; \vec 0 \, \varrho} \, \hat T_{\vec g
      + \vec d_{\mu}} \, \chi_{\mu}(\vec r) \\
    &=& \Sum_{\vec g} \Sum_{\mu=1}^K C_{\vec g + \vec R \, \mu \; \vec R \,
      \varrho} \, \chi_{\mu}(\vec r - \vec d_{\mu} - \vec g - \vec R) \; .
  \end{array}
\end{equation}
The number of basis functions~$K$ determines the number of
orbitals per unit cell.
In Eq.~(\ref{eq:Wannierbasis}), I exploit the fact that both the spatial
Wannier orbitals and the basis functions~$\hat T_{\vec g + \vec d_{\mu}}
\, \chi_{\mu}(\vec r)$ form sets of functions whose members are related
by lattice translations~(\ref{eq:WannierTrans}). The
expansion coefficients, hence, are translationally
symmetric~$C_{\vec g + \vec R \, \mu \; \vec R \, \varrho}
= C_{\vec g \, \mu \; \vec 0 \, \varrho}
\equiv C_{\mu \varrho}(\vec g)$.
The displacement~$\vec d_{\mu}$ of the $\mu$th basis function
in a unit cell accounts for the fact that a basis function
is frequently centered on atoms which, in turn, are displaced
somewhat from the origin of the unit cell.

The basis-set representation of the Fock operator~$\hat f$
is~$F_{\vec g \, \mu \; \vec g' \, \nu} =
\bra{\chi_{\vec g \, \mu}} \hat f \ket{\chi_{\vec g' \, \nu}}$,
the overlap matrix is~$S_{\vec g \, \mu \; \vec g' \, \nu} =
\bracket{\chi_{\vec g \, \mu}} {\chi_{\vec g' \, \nu}}$, and the matrix of
the penalty projection operator~(\ref{eq:penalty})---with~$N'$
replaced by~$K$---reads~\cite{Shukla:EC-96}
\begin{equation}
  \label{eq:basispen}
  \begin{array}{rcl}
    (\mat{\Cal P}_{\vec R})_{\vec g \, \mu \; \vec g' \, \nu}
    &=& \bra{\chi_{\vec g \, \mu}} \hat \Cal P_{\vec R}
    \ket{\chi_{\vec g' \, \nu}} \\
    &=& \Sum_{\vec R', \vec g_1, \vec g_2 \atop
    \vec R' \neq \vec R} \Sum_{\kappa, \xi, \varrho = 1}^K
    S  _{\vec g \, \mu \; \vec g_1 \, \kappa} \\
    &&{} \times C  _{\vec g_1 \, \kappa  \; \vec R' \, \varrho} \,
    C^*_{\vec g_2 \, \xi \; \vec R' \, \varrho} \,
    S^*_{\vec g' \, \nu \; \vec g_2 \, \xi} \; .
  \end{array}
\end{equation}
Representing Eq.~(\ref{eq:WannierFockEigenvalue}) in the basis set,
I obtain modified Roothaan-Hall equations~\cite{Shukla:EC-96}
\begin{equation}
  \label{eq:WannierRHFProj}
  (\mat F + \lambda \, \mat{\Cal P}_{\vec 0}) \, \vec C_{\varrho}
  = \varepsilon_{\vec 0 \, \varrho} \, \mat S \, \vec C_{\varrho}
\end{equation}
to which I refer as Wannier-Roothaan-Hall equations.
Formula~(\ref{eq:WannierRHFProj}) yields the occupied and virtual
Wannier orbitals in the origin cell~$\check w_{\vec 0 \, \varrho}(\vec r)$
with~$\varrho = 1, \ldots, K$ and expansion
coefficients~$(\vec C_{\varrho})_{\vec g \, \mu} \equiv C_{\vec g \, \mu
\; \vec 0 \, \varrho}$ for all~$K \, N_0$ indices~$\vec g \, \mu$.
In expression~(\ref{eq:WannierRHFProj}), the coupling of the
Wannier orbitals in the origin cell to the Wannier orbitals
in neighboring unit cells reappears by means of the basis-set
representation~(\ref{eq:Wannierbasis})
which could be avoided by omitting the Lagrangian multipliers
in Eq.~(\ref{eq:CellLagrangianMult}).
Consequently, the dimension of the matrices in Eq.~(\ref{eq:WannierRHFProj})
scales both with the number of basis functions in the origin cell
and with the number of unit cells utilized to support the
Wannier orbitals.
Note that only a subset of $K$~eigenvectors out of the $K \, N_0$~eigenvectors
of Eq.~(\ref{eq:WannierRHFProj}) is required.
The Wannier orbitals in other but the origin cell are simply
given by lattice translations~(\ref{eq:WannierTrans}), exploiting
that the expansion coefficients~$C_{\mu \varrho}(\vec g)$ are
translationally symmetric.
Therewith, all Wannier orbitals of the crystal are determined.
The energy bands are not given by the~$\varepsilon_{\vec 0 \, \varrho}$
in Eq.~(\ref{eq:WannierRHFProj}).
Instead, they are found by diagonalizing the Hermitian
matrix~$\mat F(\vec k) = \Sum_{\vec R} \euler^{\imag \vec k
\vec R} \, \mat F_{\vec 0 \, \vec R}$ for a number of $\vec k$~points,
where $(\mat F_{\vec 0 \, \vec R})_{\sigma\varrho}
= \bra{w_{\vec 0 \, \sigma}} \hat f \ket{w_{\vec R \, \varrho}}$ are
blocks of the Fock matrix in Wannier representation.

The solution of the Wannier-Roothaan-Hall equations~(\ref{eq:WannierRHFProj})
has been implemented in the \textsc{wannier}
program~\cite{Shukla:EC-96,Shukla:WF-98} and its accuracy was
tested in a series of studies.
Systematic comparisons were made of ground-state properties
and band structures of crystals~\cite{Shukla:WF-98,Albrecht:HF-98} and
infinite chains.~\cite{Abdurahman:AI-99}
The basis-set expansion~(\ref{eq:Wannierbasis}) is found
to provide a satisfactory accuracy using up to
third-nearest-neighbor cells for ionic three-dimensional
crystals~\cite{Shukla:EC-96,Shukla:WF-98,Albrecht:HF-98,Shukla:TC-98,%
Buth:MC-05,Buth:AG-05}
as well as covalently bonded and hydrogen-bonded infinite
chains.~\cite{Abdurahman:AI-99,Buth:BS-04,Buth:MC-05,Buth:HB-06,Buth:BHF-06}
Note, however, that despite the fact that variation of the expansion coefficients
in Eq.~(\ref{eq:Wannierbasis}) is allowed only in a small cluster of unit
cells, a sophisticated treatment of the very long range electrostatic interactions
in the Fock operator is required, accounting for a large number of unit cells.
Therefore, one uses the Ewald summation technique for the Coulomb matrix
elements and an extended summation for the exchange matrix
elements.~\cite{Shukla:WF-98}
The occupied pseudocanonical \emph{a priori} Wannier orbitals are
found to reach a few angstroms from the atom they are
centered on.~\cite{Shukla:WF-98,Shukla:TC-98}

\section{Properties}
\label{sec:properties}
\subsection{Relation to canonical Hartree-Fock equations}

The connection of the Wannier-Hartree-Fock
equations~(\ref{eq:WannierFockEigenvalue}) to the corresponding
equations of a Bloch-orbital-based formalism---the canonical Hartree-Fock
equations~\cite{Szabo:MQC-89,McWeeny:MQM-92}---can be achieved readily.
To this end, the orthogonalizing potential~(\ref{eq:orthogpot})
is replaced by an expression which comprises the full overlap matrix, \ie,
$\frac{\lambda}{2} \, \Tr \mat S^2$.
As the Wannier functions are normalized to unity, one
obtains~$\Tr \mat S^2 = \Tr \mat \offS^2 + K \, N_0$.
This amounts to a meaningless overall energetic shift of~$\Cal L'$
in Eq.~(\ref{eq:WannierRitz}) upon replacing~$V_{\rm Orth}$.
Functional variation of the new~$\Cal L'$ yields that
the summation in the penalty projector of Eqs.~(\ref{eq:penalty})
and (\ref{eq:basispen}) is no longer restricted to the unit cells in the
neighborhood but also contains the origin cell, \ie, the
operator~$\Sum_{\vec R' \neq \vec R}$ is exchanged by~$\Sum_{\vec R'}$.
The new projector~$\hat{\Cal P}$ becomes translationally
symmetric and actually the identity operator~$\unitop$.
This transformation of the Hartree-Fock equations is equivalent to the
addition of~$\Sum_{\xi=1}^{N^\prime} \ket{w_{\vec R \, \xi}}
\bra{w_{\vec R \, \xi}}$ on both sides of
Eq.~(\ref{eq:WannierFockPenalty}), which causes the
eigenvalues~$\varepsilon_{\vec 0 \, \kappa}$
[cf.~Eq.~(\ref{eq:WannierFockEigenvalue})] to shift by~$\lambda$,
\ie, they become~$\varepsilon_{\vec 0 \, \kappa} + \lambda$.
The modification undoes the orthogonalizing potential and
reverts the equations to the modified Hartree-Fock equations
which result from Eq.~(\ref{eq:CellLagrangianMult})
with its eigenvalues shifted by~$\lambda$.
Expression~(\ref{eq:WannierFock}) is obtained again by realizing that
the matrix representation of~$\hat f + \lambda \, \hat{\Cal P}$ in terms
of Wannier orbitals also contains off-diagonal terms with respect to the
lattice vectors, thus reintroducing off-diagonal Lagrangian multipliers.

The arguments of the previous paragraph can be expressed
more clearly by changing the Roothaan-Hall
equations~(\ref{eq:WannierRHFProj}) because their dimensionality and formal
structure are preserved under the replacement of the projector.
Due to the translational symmetry of~$\hat f + \hat{\Cal P}$,
Born-von K\'arm\'an boundary conditions become
beneficial.~\cite{Ashcroft:SSP-76,Callaway:QT-91}
With the matrix representation of~$\hat{\Cal P} = \unitop$, which is
$\mat{\Cal P} = \mat S$, the modified Roothaan-Hall
equations~(\ref{eq:WannierRHFProj}) are changed to~\cite{Buth:MC-05}
\begin{equation}
  \label{eq:transRoothaanHall}
  (\mat F + \lambda \, \mat{\Cal P}) \, \mat C'
    = \mat S \, \mat C' \, \mat \varepsilon' \; .
\end{equation}
The $N_0 \, K \times N_0 \, K$~matrices~$\mat F$, $\mat S$, and
$\mat{\Cal P}$ are cyclic matrices which can be block diagonalized
employing the unitary transformation~\cite{Lowdin:QT-56,%
Ladik:PS-99}~$\Cal W_{\vec R \, \varrho \; \vec k \, p}
= \frac{1}{\sqrt{N_0}} \, \delta_{\varrho \, p} \,
\euler^{\imag \vec k \, \vec R}$.
Multiplying with~$\mat{\Cal W}^{\dagger}$ from the left
and inserting~$\mat{\Cal W} \mat{\Cal W}^{\dagger}$ before~$\mat C'$,
the eigenvalue problem~(\ref{eq:transRoothaanHall}) breaks down into
$N_0$~independent $K \times K$~eigenvalue problems,
\begin{equation}
  \label{eq:WRHFBvK}
  [\mat F(\vec k) + \lambda \, \mat{\Cal P}(\vec k)] \, \mat C'(\vec k)
  = \mat S(\vec k) \, \mat C'(\vec k) \, \mat\varepsilon'(\vec k) \; .
\end{equation}
Solving these equations yields orthonormal Bloch orbitals and energy
bands which are shifted by~$\lambda$.

\subsection{Computational efficiency}

To compare the computational effort of the canonical Hartree-Fock
method to the modified Hartree-Fock method, I analyze the
corresponding Roothaan-Hall equations, Eqs.~(\ref{eq:WRHFBvK}) and
(\ref{eq:WannierRHFProj}), respectively.
Without regarding space-group symmetry, both sets of equations
have the same dimensionality because the number of orbitals (or basis
functions) is the same, namely, $K \, N_0$.
The effort to compute the matrix elements is clearly the same
for the basis-set overlap matrix;
it is also equal for the Fock matrix because the Fock operator
is invariant under unitary transformations of the orbitals.
The support of the basis-set expansion is the same for Bloch and
Wannier orbitals:
the basis functions in the parallelepiped of $N_0$~unit cells.
Only the expansion coefficients differ.
Consequently, the number of floating point operations necessary to determine
the Fock operator is the same, if no further approximations are made.
Yet the projection operator, which is only required for the modified
Hartree-Fock equations, requires an extra effort.

In a next step, the modified and canonical Roothaan-Hall equations
need to be diagonalized.
In both cases, a full diagonalization requires the same effort.
However, in practice a selective computation is carried out.
The canonical equations are block diagonalized first, Eq.~(\ref{eq:WRHFBvK}),
and the subblocks are independently diagonalized fully afterwards
for a grid of $\vec k$~points.~\cite{Monkhorst:SP-76}
The Wannier-Roothaan-Hall equations have to be treated differently.
The spectrum of~$\mat F + \lambda \mat{\Cal P}_{\vec 0}$
in Eq.~(\ref{eq:WannierRHFProj}) has a very favorable property:
the lower $K$~eigenvalues correspond to
the Wannier orbitals in the origin cell.
The other eigenvalues are well separated from the former ones
because they are shifted to high values by the shift parameter~$\lambda$.~\cite{Shukla:WF-98}
Iterative eigenvalue solvers, particularly
the one of Davidson,~\cite{Davidson:IC-75}
can be employed to reduce the numerical effort
to determine the lowest $K$~eigenpairs
of~$\mat F + \lambda \mat{\Cal P}_{\vec 0}$.
Their effort is predominantly determined by the matrix times vector product.
As the summation expression for this product and the summations
for the block diagonalization and subsequent $\vec k$-space integration
are similar, the overall computational effort of both methods, to determine
crystal orbitals with a certain accuracy, should be comparable.
In fact, the Bloch-orbital-based equations can be solved more efficiently
for crystals with a small number of atoms per unit cell;
the solution of the Wannier-orbital-based equations is more efficient for
crystals with a large unit cell because cutoff criteria can be established
to lower the actual effort.

\section{Conclusion}
\label{sec:conclusion}

In this paper, I derive modified Hartree-Fock equations,
which directly yield Wannier orbitals.
An orthogonalizing potential is added to the Hartree-Fock energy
functional to ensure the proper orthonormality of the resulting orbitals.
It serves to replace the Lagrangian multipliers between unit cells
needed otherwise.
The equations necessarily break the translational symmetry as Wannier
functions are translationally related, in contrast to Bloch functions
which are translationally symmetric.
I show how the conventional Bloch-orbital-based
Hartree-Fock equations can be recovered by restoring the translational
symmetry of the modified equations.
Analyzing the spectral properties of the modified Fock
matrix, I find the numerical effort of the method
to be comparable to the effort of the Bloch-orbital-based approach.

The orbitals which result from the modified Hartree-Fock equations
are referred to as pseudocanonical Wannier functions as
they are delocalized over the entire unit cell.
However, they can be localized additionally within unit cells by adding a
suitable localizing potential to the energy expression.
Particularly, the potential of Gilbert~\cite{Gilbert:SC-64}
is to be mentioned here, which minimizes the Foster-Boys
functional.~\cite{Foster:MO-60,Boys:MO-60,Boys:-66}
It thus offers a different route to
Refs.~\onlinecite{Marzari:ML-97,Zicovich:GM-01}
to determine maximally localized Wannier functions.
Alternatively, the potential of Edmiston and Ruedenberg
can be used as a localizing potential.~\cite{Edmiston:LA-63}

This study is based on the Hartree-Fock theory.
However, the extension of the ideas to density-functional
theory~\cite{Hohenberg:IEG-64,Kohn:SC-65,Parr:DF-89} is straightforward.
To this end, one exchanges the Hartree-Fock energy
functional~(\ref{eq:WannierFockEnergy}) by the Hohenberg-Kohn energy
functional in terms of the Kohn-Sham orbitals.~\cite{Kohn:SC-65,Parr:DF-89}
The Hohenberg-Kohn variational theorem~\cite{Hohenberg:IEG-64,Parr:DF-89}
ensures that minimizing this functional with respect to the orbitals
provides the exact ground-state energy, if one uses the exact
exchange-correlation energy functional.
The minimization is carried out under the constraint that the orbitals
remain orthonormal using Eq.~(\ref{eq:LagrangianMult}).
In other words, we can essentially follow the line of argument
that leads from Eq.~(\ref{eq:WannierFockEnergy}) to
Eq.~(\ref{eq:WannierFockEigenvalue}).
I term the latter equation---with the Fock operator replaced by
the Kohn-Sham operator---pseudocanonical Wannier-Kohn-Sham
equations.
All subsequent arguments and expressions carry over analogously.

\begin{acknowledgments}
I would like to thank Hendrik J.~Monkhorst and Robin Santra for helpful
discussions and critical reading of the paper.
Furthermore, my gratitude is extended to Martin Albrecht for
introducing me to the \textsc{wannier} program.
This work was partly supported by a Feodor Lynen Research Fellowship
from the Alexander von Humboldt Foundation.
\end{acknowledgments}


\begin{thebibliography}{43}
\expandafter\ifx\csname natexlab\endcsname\relax\def\natexlab#1{#1}\fi
\expandafter\ifx\csname bibnamefont\endcsname\relax
  \def\bibnamefont#1{#1}\fi
\expandafter\ifx\csname bibfnamefont\endcsname\relax
  \def\bibfnamefont#1{#1}\fi
\expandafter\ifx\csname citenamefont\endcsname\relax
  \def\citenamefont#1{#1}\fi
\expandafter\ifx\csname url\endcsname\relax
  \def\url#1{\texttt{#1}}\fi
\expandafter\ifx\csname urlprefix\endcsname\relax\def\urlprefix{URL }\fi
\providecommand{\bibinfo}[2]{#2}
\providecommand{\eprint}[2][]{\url{#2}}

\bibitem[{\citenamefont{Wannier}(1937)}]{Wannier:EE-37}
\bibinfo{author}{\bibfnamefont{G.~H.} \bibnamefont{Wannier}},
  \bibinfo{journal}{Phys. Rev.} \textbf{\bibinfo{volume}{52}},
  \bibinfo{pages}{191} (\bibinfo{year}{1937}).

\bibitem[{\citenamefont{Blount}(1962)}]{Blount:FB-62}
\bibinfo{author}{\bibfnamefont{E.~I.} \bibnamefont{Blount}}, in
  \emph{\bibinfo{booktitle}{Solid state physics}}, edited by
  \bibinfo{editor}{\bibfnamefont{F.}~\bibnamefont{Seitz}} \bibnamefont{and}
  \bibinfo{editor}{\bibfnamefont{D.}~\bibnamefont{Turnbull}}
  (\bibinfo{publisher}{Academic Press}, \bibinfo{address}{New York},
  \bibinfo{year}{1962}), vol.~\bibinfo{volume}{13}, pp.
  \bibinfo{pages}{305--373}.

\bibitem[{\citenamefont{Ashcroft and Mermin}(1976)}]{Ashcroft:SSP-76}
\bibinfo{author}{\bibfnamefont{N.~W.} \bibnamefont{Ashcroft}} \bibnamefont{and}
  \bibinfo{author}{\bibfnamefont{N.~D.} \bibnamefont{Mermin}},
  \emph{\bibinfo{title}{Solid state physics}} (\bibinfo{publisher}{Cole},
  \bibinfo{address}{London}, \bibinfo{year}{1976}), ISBN
  \bibinfo{isbn}{0-03-083993-9}.

\bibitem[{\citenamefont{Callaway}(1991)}]{Callaway:QT-91}
\bibinfo{author}{\bibfnamefont{J.}~\bibnamefont{Callaway}},
  \emph{\bibinfo{title}{Quantum theory of the solid state}}
  (\bibinfo{publisher}{Academic Press}, \bibinfo{address}{Boston},
  \bibinfo{year}{1991}), \bibinfo{edition}{2nd} ed., ISBN
  \bibinfo{isbn}{0-12-155203-9}.

\bibitem[{\citenamefont{Fulde}(1995)}]{Fulde:EC-95}
\bibinfo{author}{\bibfnamefont{P.}~\bibnamefont{Fulde}},
  \emph{\bibinfo{title}{Electron correlations in molecules and solids}}, vol.
  \bibinfo{volume}{100} of \emph{\bibinfo{series}{Springer series in
  solid-state sciences}} (\bibinfo{publisher}{Springer},
  \bibinfo{address}{Berlin}, \bibinfo{year}{1995}), \bibinfo{edition}{3rd} ed.,
  ISBN \bibinfo{isbn}{3-540-59364-0}.

\bibitem[{\citenamefont{Marzari and Vanderbilt}(1997)}]{Marzari:ML-97}
\bibinfo{author}{\bibfnamefont{N.}~\bibnamefont{Marzari}} \bibnamefont{and}
  \bibinfo{author}{\bibfnamefont{D.}~\bibnamefont{Vanderbilt}},
  \bibinfo{journal}{Phys. Rev.~B} \textbf{\bibinfo{volume}{56}},
  \bibinfo{pages}{12847} (\bibinfo{year}{1997}).

\bibitem[{\citenamefont{Zicovich-Wilson
  et~al.}(2001)\citenamefont{Zicovich-Wilson, Dovesi, and
  Saunders}}]{Zicovich:GM-01}
\bibinfo{author}{\bibfnamefont{C.~M.} \bibnamefont{Zicovich-Wilson}},
  \bibinfo{author}{\bibfnamefont{R.}~\bibnamefont{Dovesi}}, \bibnamefont{and}
  \bibinfo{author}{\bibfnamefont{V.~R.} \bibnamefont{Saunders}},
  \bibinfo{journal}{J. Chem. Phys.} \textbf{\bibinfo{volume}{115}},
  \bibinfo{pages}{9708} (\bibinfo{year}{2001}).

\bibitem[{\citenamefont{Fulde}(2002)}]{Fulde:WF-02}
\bibinfo{author}{\bibfnamefont{P.}~\bibnamefont{Fulde}}, \bibinfo{journal}{Adv.
  Phys.} \textbf{\bibinfo{volume}{51}}, \bibinfo{pages}{909}
  (\bibinfo{year}{2002}).

\bibitem[{\citenamefont{Goedecker}(1999)}]{Goedecker:LS-99}
\bibinfo{author}{\bibfnamefont{S.}~\bibnamefont{Goedecker}},
  \bibinfo{journal}{Rev. Mod. Phys.} \textbf{\bibinfo{volume}{71}},
  \bibinfo{pages}{1085} (\bibinfo{year}{1999}).

\bibitem[{\citenamefont{Wu and Jayanthi}(2002)}]{Wu:ON-02}
\bibinfo{author}{\bibfnamefont{S.~Y.} \bibnamefont{Wu}} \bibnamefont{and}
  \bibinfo{author}{\bibfnamefont{C.~S.} \bibnamefont{Jayanthi}},
  \bibinfo{journal}{Phys. Rep.} \textbf{\bibinfo{volume}{358}},
  \bibinfo{pages}{1} (\bibinfo{year}{2002}).

\bibitem[{\citenamefont{Shukla et~al.}(1999)\citenamefont{Shukla, Dolg, Fulde,
  and Stoll}}]{Shukla:WFB-99}
\bibinfo{author}{\bibfnamefont{A.}~\bibnamefont{Shukla}},
  \bibinfo{author}{\bibfnamefont{M.}~\bibnamefont{Dolg}},
  \bibinfo{author}{\bibfnamefont{P.}~\bibnamefont{Fulde}}, \bibnamefont{and}
  \bibinfo{author}{\bibfnamefont{H.}~\bibnamefont{Stoll}},
  \bibinfo{journal}{Phys. Rev.~B} \textbf{\bibinfo{volume}{60}},
  \bibinfo{pages}{5211} (\bibinfo{year}{1999}).

\bibitem[{\citenamefont{Albrecht and Fulde}(2002)}]{Albrecht:LAI-02}
\bibinfo{author}{\bibfnamefont{M.}~\bibnamefont{Albrecht}} \bibnamefont{and}
  \bibinfo{author}{\bibfnamefont{P.}~\bibnamefont{Fulde}},
  \bibinfo{journal}{phys. stat. sol. (b)} \textbf{\bibinfo{volume}{234}},
  \bibinfo{pages}{313} (\bibinfo{year}{2002}).

\bibitem[{\citenamefont{Buth and Paulus}(2004)}]{Buth:BS-04}
\bibinfo{author}{\bibfnamefont{C.}~\bibnamefont{Buth}} \bibnamefont{and}
  \bibinfo{author}{\bibfnamefont{B.}~\bibnamefont{Paulus}},
  \bibinfo{journal}{Chem. Phys. Lett.} \textbf{\bibinfo{volume}{398}},
  \bibinfo{pages}{44} (\bibinfo{year}{2004}),
  \bibinfo{note}{\href{http://de.arxiv.org/abs/cond-mat/0408243}
  {arXiv:cond-mat/0408243}}.

\bibitem[{\citenamefont{Buth}(2005)}]{Buth:MC-05}
\bibinfo{author}{\bibfnamefont{C.}~\bibnamefont{Buth}},
  \bibinfo{type}{Dissertation}, \bibinfo{school}{Technische Universit\"at
  Dresden}, \bibinfo{address}{01062~Dresden, Germany} (\bibinfo{year}{2005}),
  \bibinfo{note}{\href{http://nbn-resolving.de/urn:nbn:de:swb:14-1132580113554%
-34509} {nbn-resolving.de/urn:nbn:de:swb:14-1132580113554-34509}}.

\bibitem[{\citenamefont{Buth et~al.}(2005)\citenamefont{Buth, Birkenheuer,
  Albrecht, and Fulde}}]{Buth:AG-05}
\bibinfo{author}{\bibfnamefont{C.}~\bibnamefont{Buth}},
  \bibinfo{author}{\bibfnamefont{U.}~\bibnamefont{Birkenheuer}},
  \bibinfo{author}{\bibfnamefont{M.}~\bibnamefont{Albrecht}}, \bibnamefont{and}
  \bibinfo{author}{\bibfnamefont{P.}~\bibnamefont{Fulde}},
  \bibinfo{journal}{Phys. Rev. B} \textbf{\bibinfo{volume}{72}},
  \bibinfo{pages}{195107} (\bibinfo{year}{2005}),
  \bibinfo{note}{\href{http://de.arxiv.org/abs/cond-mat/0409078}
  {arXiv:cond-mat/0409078}}.

\bibitem[{\citenamefont{Buth and Paulus}(2006)}]{Buth:HB-06}
\bibinfo{author}{\bibfnamefont{C.}~\bibnamefont{Buth}} \bibnamefont{and}
  \bibinfo{author}{\bibfnamefont{B.}~\bibnamefont{Paulus}},
  \bibinfo{journal}{Phys. Rev.~B} \textbf{\bibinfo{volume}{74}},
  \bibinfo{pages}{045122} (\bibinfo{year}{2006}),
  \bibinfo{note}{\href{http://de.arxiv.org/abs/cond-mat/0601470}
  {arXiv:cond-mat/0601470}}.

\bibitem[{\citenamefont{Buth}(2006)}]{Buth:BHF-06}
\bibinfo{author}{\bibfnamefont{C.}~\bibnamefont{Buth}}, \bibinfo{journal}{J.
  Chem. Phys.} \textbf{\bibinfo{volume}{125}}, \bibinfo{pages}{154707}
  (\bibinfo{year}{2006}),
  \bibinfo{note}{\href{http://de.arxiv.org/abs/cond-mat/0606081}
  {arXiv:cond-mat/0606081}}.

\bibitem[{\citenamefont{Pisani et~al.}(2005)\citenamefont{Pisani, Busso,
  Capecchi, Casassa, Dovesi, Maschio, Zicovich-Wilson, and
  Sch\"utz}}]{Pisani:MP-06}
\bibinfo{author}{\bibfnamefont{C.}~\bibnamefont{Pisani}},
  \bibinfo{author}{\bibfnamefont{M.}~\bibnamefont{Busso}},
  \bibinfo{author}{\bibfnamefont{G.}~\bibnamefont{Capecchi}},
  \bibinfo{author}{\bibfnamefont{S.}~\bibnamefont{Casassa}},
  \bibinfo{author}{\bibfnamefont{R.}~\bibnamefont{Dovesi}},
  \bibinfo{author}{\bibfnamefont{L.}~\bibnamefont{Maschio}},
  \bibinfo{author}{\bibfnamefont{C.}~\bibnamefont{Zicovich-Wilson}},
  \bibnamefont{and} \bibinfo{author}{\bibfnamefont{M.}~\bibnamefont{Sch\"utz}},
  \bibinfo{journal}{J. Chem. Phys.} \textbf{\bibinfo{volume}{122}},
  \bibinfo{pages}{094113} (\bibinfo{year}{2005}).

\bibitem[{\citenamefont{Foster and Boys}(1960)}]{Foster:MO-60}
\bibinfo{author}{\bibfnamefont{J.~M.} \bibnamefont{Foster}} \bibnamefont{and}
  \bibinfo{author}{\bibfnamefont{S.~F.} \bibnamefont{Boys}},
  \bibinfo{journal}{Rev. Mod. Phys.} \textbf{\bibinfo{volume}{32}},
  \bibinfo{pages}{300} (\bibinfo{year}{1960}).

\bibitem[{\citenamefont{Boys}(1960)}]{Boys:MO-60}
\bibinfo{author}{\bibfnamefont{S.~F.} \bibnamefont{Boys}},
  \bibinfo{journal}{Rev. Mod. Phys.} \textbf{\bibinfo{volume}{32}},
  \bibinfo{pages}{296} (\bibinfo{year}{1960}).

\bibitem[{\citenamefont{Boys}(1966)}]{Boys:-66}
\bibinfo{author}{\bibfnamefont{S.~F.} \bibnamefont{Boys}}, in
  \emph{\bibinfo{booktitle}{Quantum theory of atoms, molecules, and the solid
  state}}, edited by \bibinfo{editor}{\bibfnamefont{P.-O.}
  \bibnamefont{L\"owdin}} (\bibinfo{publisher}{Academic Press},
  \bibinfo{address}{New York}, \bibinfo{year}{1966}), p. \bibinfo{pages}{253}.

\bibitem[{\citenamefont{Edmiston and Ruedenberg}(1963)}]{Edmiston:LA-63}
\bibinfo{author}{\bibfnamefont{C.}~\bibnamefont{Edmiston}} \bibnamefont{and}
  \bibinfo{author}{\bibfnamefont{K.}~\bibnamefont{Ruedenberg}},
  \bibinfo{journal}{Rev. Mod. Phys.} \textbf{\bibinfo{volume}{35}},
  \bibinfo{pages}{457} (\bibinfo{year}{1963}).

\bibitem[{\citenamefont{Pipek and Mezey}(1989)}]{Pipek:FI-89}
\bibinfo{author}{\bibfnamefont{J.}~\bibnamefont{Pipek}} \bibnamefont{and}
  \bibinfo{author}{\bibfnamefont{P.~G.} \bibnamefont{Mezey}},
  \bibinfo{journal}{J. Chem. Phys.} \textbf{\bibinfo{volume}{90}},
  \bibinfo{pages}{4916} (\bibinfo{year}{1989}).

\bibitem[{\citenamefont{Shukla et~al.}(1996)\citenamefont{Shukla, Dolg, Stoll,
  and Fulde}}]{Shukla:EC-96}
\bibinfo{author}{\bibfnamefont{A.}~\bibnamefont{Shukla}},
  \bibinfo{author}{\bibfnamefont{M.}~\bibnamefont{Dolg}},
  \bibinfo{author}{\bibfnamefont{H.}~\bibnamefont{Stoll}}, \bibnamefont{and}
  \bibinfo{author}{\bibfnamefont{P.}~\bibnamefont{Fulde}},
  \bibinfo{journal}{Chem. Phys. Lett.} \textbf{\bibinfo{volume}{262}},
  \bibinfo{pages}{213} (\bibinfo{year}{1996}).

\bibitem[{\citenamefont{Shukla et~al.}(1998{\natexlab{a}})\citenamefont{Shukla,
  Dolg, Fulde, and Stoll}}]{Shukla:WF-98}
\bibinfo{author}{\bibfnamefont{A.}~\bibnamefont{Shukla}},
  \bibinfo{author}{\bibfnamefont{M.}~\bibnamefont{Dolg}},
  \bibinfo{author}{\bibfnamefont{P.}~\bibnamefont{Fulde}}, \bibnamefont{and}
  \bibinfo{author}{\bibfnamefont{H.}~\bibnamefont{Stoll}},
  \bibinfo{journal}{Phys. Rev. B} \textbf{\bibinfo{volume}{57}},
  \bibinfo{pages}{1471} (\bibinfo{year}{1998}{\natexlab{a}}).

\bibitem[{\citenamefont{Shukla et~al.}(1998{\natexlab{b}})\citenamefont{Shukla,
  Dolg, Fulde, and Stoll}}]{Shukla:TC-98}
\bibinfo{author}{\bibfnamefont{A.}~\bibnamefont{Shukla}},
  \bibinfo{author}{\bibfnamefont{M.}~\bibnamefont{Dolg}},
  \bibinfo{author}{\bibfnamefont{P.}~\bibnamefont{Fulde}}, \bibnamefont{and}
  \bibinfo{author}{\bibfnamefont{H.}~\bibnamefont{Stoll}}, \bibinfo{journal}{J.
  Chem. Phys.} \textbf{\bibinfo{volume}{108}}, \bibinfo{pages}{8521}
  (\bibinfo{year}{1998}{\natexlab{b}}).

\bibitem[{\citenamefont{Albrecht et~al.}(1998)\citenamefont{Albrecht, Shukla,
  Dolg, Fulde, and Stoll}}]{Albrecht:HF-98}
\bibinfo{author}{\bibfnamefont{M.}~\bibnamefont{Albrecht}},
  \bibinfo{author}{\bibfnamefont{A.}~\bibnamefont{Shukla}},
  \bibinfo{author}{\bibfnamefont{M.}~\bibnamefont{Dolg}},
  \bibinfo{author}{\bibfnamefont{P.}~\bibnamefont{Fulde}}, \bibnamefont{and}
  \bibinfo{author}{\bibfnamefont{H.}~\bibnamefont{Stoll}},
  \bibinfo{journal}{Chem. Phys. Lett.} \textbf{\bibinfo{volume}{285}},
  \bibinfo{pages}{174} (\bibinfo{year}{1998}).

\bibitem[{\citenamefont{Abdurahman et~al.}(1999)\citenamefont{Abdurahman,
  Albrecht, Shukla, and Dolg}}]{Abdurahman:AI-99}
\bibinfo{author}{\bibfnamefont{A.}~\bibnamefont{Abdurahman}},
  \bibinfo{author}{\bibfnamefont{M.}~\bibnamefont{Albrecht}},
  \bibinfo{author}{\bibfnamefont{A.}~\bibnamefont{Shukla}}, \bibnamefont{and}
  \bibinfo{author}{\bibfnamefont{M.}~\bibnamefont{Dolg}}, \bibinfo{journal}{J.
  Chem. Phys.} \textbf{\bibinfo{volume}{110}}, \bibinfo{pages}{8819}
  (\bibinfo{year}{1999}).

\bibitem[{\citenamefont{Abdurahman et~al.}(2000)\citenamefont{Abdurahman,
  Shukla, and Dolg}}]{Abdurahman:AI-00}
\bibinfo{author}{\bibfnamefont{A.}~\bibnamefont{Abdurahman}},
  \bibinfo{author}{\bibfnamefont{A.}~\bibnamefont{Shukla}}, \bibnamefont{and}
  \bibinfo{author}{\bibfnamefont{M.}~\bibnamefont{Dolg}}, \bibinfo{journal}{J.
  Chem. Phys.} \textbf{\bibinfo{volume}{112}}, \bibinfo{pages}{4801}
  (\bibinfo{year}{2000}).

\bibitem[{\citenamefont{Mauri and Galli}(1994)}]{Mauri:EC-94}
\bibinfo{author}{\bibfnamefont{F.}~\bibnamefont{Mauri}} \bibnamefont{and}
  \bibinfo{author}{\bibfnamefont{G.}~\bibnamefont{Galli}},
  \bibinfo{journal}{Phys. Rev.~B} \textbf{\bibinfo{volume}{50}},
  \bibinfo{pages}{4316} (\bibinfo{year}{1994}).

\bibitem[{\citenamefont{Ladik}(1999)}]{Ladik:PS-99}
\bibinfo{author}{\bibfnamefont{J.~J.} \bibnamefont{Ladik}},
  \bibinfo{journal}{Phys. Rep.} \textbf{\bibinfo{volume}{313}},
  \bibinfo{pages}{171} (\bibinfo{year}{1999}).

\bibitem[{\citenamefont{McWeeny}(1992)}]{McWeeny:MQM-92}
\bibinfo{author}{\bibfnamefont{R.}~\bibnamefont{McWeeny}},
  \emph{\bibinfo{title}{Methods of molecular quantum mechanics}}
  (\bibinfo{publisher}{Academic Press}, \bibinfo{address}{London},
  \bibinfo{year}{1992}), \bibinfo{edition}{2nd} ed., ISBN
  \bibinfo{isbn}{0-12-486551-8}.

\bibitem[{\citenamefont{Szabo and Ostlund}(1989)}]{Szabo:MQC-89}
\bibinfo{author}{\bibfnamefont{A.}~\bibnamefont{Szabo}} \bibnamefont{and}
  \bibinfo{author}{\bibfnamefont{N.~S.} \bibnamefont{Ostlund}},
  \emph{\bibinfo{title}{Modern quantum chemistry: Introduction to advanced
  electronic structure theory}} (\bibinfo{publisher}{McGraw-Hill},
  \bibinfo{address}{New York}, \bibinfo{year}{1989}), \bibinfo{edition}{{1st,
  revised}} ed., ISBN \bibinfo{isbn}{0-486-69186-1}.

\bibitem[{\citenamefont{Del~Re et~al.}(1967)\citenamefont{Del~Re, Ladik, and
  Bicz\'o}}]{Re:SC-67}
\bibinfo{author}{\bibfnamefont{G.}~\bibnamefont{Del~Re}},
  \bibinfo{author}{\bibfnamefont{J.~J.} \bibnamefont{Ladik}}, \bibnamefont{and}
  \bibinfo{author}{\bibfnamefont{G.}~\bibnamefont{Bicz\'o}},
  \bibinfo{journal}{Phys. Rev.} \textbf{\bibinfo{volume}{155}},
  \bibinfo{pages}{997} (\bibinfo{year}{1967}).

\bibitem[{\citenamefont{Andr\'e et~al.}(1967)\citenamefont{Andr\'e, Gouverneur,
  and Leroy}}]{Andre:ET-67}
\bibinfo{author}{\bibfnamefont{J.-M.} \bibnamefont{Andr\'e}},
  \bibinfo{author}{\bibfnamefont{L.}~\bibnamefont{Gouverneur}},
  \bibnamefont{and} \bibinfo{author}{\bibfnamefont{G.}~\bibnamefont{Leroy}},
  \bibinfo{journal}{Int. J. Quantum Chem.} \textbf{\bibinfo{volume}{1}},
  \bibinfo{pages}{451} (\bibinfo{year}{1967}).

\bibitem[{\citenamefont{Gilbert}(1964)}]{Gilbert:SC-64}
\bibinfo{author}{\bibfnamefont{T.~L.} \bibnamefont{Gilbert}}, in
  \emph{\bibinfo{booktitle}{Molecular orbitals in chemistry, physics and
  biology}}, edited by \bibinfo{editor}{\bibfnamefont{P.-O.}
  \bibnamefont{L\"owdin}} \bibnamefont{and}
  \bibinfo{editor}{\bibfnamefont{B.}~\bibnamefont{Pullman}}
  (\bibinfo{publisher}{Academic Press}, \bibinfo{address}{New York},
  \bibinfo{year}{1964}), pp. \bibinfo{pages}{405--420}.

\bibitem[{\citenamefont{L\"owdin}(1956)}]{Lowdin:QT-56}
\bibinfo{author}{\bibfnamefont{P.-O.} \bibnamefont{L\"owdin}},
  \bibinfo{journal}{Adv. Phys.} \textbf{\bibinfo{volume}{5}},
  \bibinfo{pages}{1} (\bibinfo{year}{1956}).

\bibitem[{\citenamefont{Monkhorst and Pack}(1976)}]{Monkhorst:SP-76}
\bibinfo{author}{\bibfnamefont{H.~J.} \bibnamefont{Monkhorst}}
  \bibnamefont{and} \bibinfo{author}{\bibfnamefont{J.~D.} \bibnamefont{Pack}},
  \bibinfo{journal}{Phys. Rev.~B} \textbf{\bibinfo{volume}{13}},
  \bibinfo{pages}{5188} (\bibinfo{year}{1976}).

\bibitem[{\citenamefont{Davidson}(1975)}]{Davidson:IC-75}
\bibinfo{author}{\bibfnamefont{E.~R.} \bibnamefont{Davidson}},
  \bibinfo{journal}{J. Comp. Phys.} \textbf{\bibinfo{volume}{17}},
  \bibinfo{pages}{87} (\bibinfo{year}{1975}).

\bibitem[{\citenamefont{Hohenberg and Kohn}(1964)}]{Hohenberg:IEG-64}
\bibinfo{author}{\bibfnamefont{P.}~\bibnamefont{Hohenberg}} \bibnamefont{and}
  \bibinfo{author}{\bibfnamefont{W.}~\bibnamefont{Kohn}},
  \bibinfo{journal}{Phys. Rev.} \textbf{\bibinfo{volume}{136}},
  \bibinfo{pages}{B864} (\bibinfo{year}{1964}).

\bibitem[{\citenamefont{Kohn and Sham}(1965)}]{Kohn:SC-65}
\bibinfo{author}{\bibfnamefont{W.}~\bibnamefont{Kohn}} \bibnamefont{and}
  \bibinfo{author}{\bibfnamefont{L.~J.} \bibnamefont{Sham}},
  \bibinfo{journal}{Phys. Rev.} \textbf{\bibinfo{volume}{140}},
  \bibinfo{pages}{A1133} (\bibinfo{year}{1965}).

\bibitem[{\citenamefont{Parr and Yang}(1989)}]{Parr:DF-89}
\bibinfo{author}{\bibfnamefont{R.~G.} \bibnamefont{Parr}} \bibnamefont{and}
  \bibinfo{author}{\bibfnamefont{W.}~\bibnamefont{Yang}},
  \emph{\bibinfo{title}{Density-functional theory of atoms and molecules}},
  vol.~\bibinfo{volume}{16} of \emph{\bibinfo{series}{International series of
  monographs on chemstry}} (\bibinfo{publisher}{Oxford University Press},
  \bibinfo{address}{Oxford, New York}, \bibinfo{year}{1989}), ISBN
  \bibinfo{isbn}{0-19-504279-4}.

\bibitem[{\citenamefont{Adams}(1961)}]{Adams:SH-61}
\bibinfo{author}{\bibfnamefont{W.~H.} \bibnamefont{Adams}},
  \bibinfo{journal}{J. Chem. Phys.} \textbf{\bibinfo{volume}{34}},
  \bibinfo{pages}{89} (\bibinfo{year}{1961}).
\end{thebibliography}
\end{document}